\documentclass[twocolumn,showpacs,preprintnumbers,amsmath,amssymb]{revtex4}

\usepackage{graphicx}
\usepackage{dcolumn}
\usepackage{bm}

\begin{document}

\title{Electronic Properties of the Semiconductor RuIn$_3$}

\author{D. Bogdanov}
 \email{dbogdan@gwdg.de}
\author{K. Winzer}

\affiliation{
1.Physikalisches Institut, Georg-August-Universit\"at G\"ottingen, D-37077 G\"ottingen, Germany}

\author{I.A. Nekrasov}
\affiliation{
Institute of Electrophysics, Russian Academy of Sciences, 620016, Ekaterinburg, Russia}

\author{T. Pruschke}
\affiliation{
Institut f\"ur Theoretische Physik, Georg-August-Universit\"at G\"ottingen, D-37077 G\"ottingen, Germany}

\date{\today}

\begin{abstract}
Temperature dependent measurements of the resistivity on RuIn$_3$
single crystals show a semiconducting behaviour, in contrast to
previously published results \cite{Roof,Poettgen}. In the high
temperature range the semiconducting gap was measured to be $0.4-0.5$
eV. We observe an anisotropy of the resistivity along [110] and [001]
orientations of the tetragonal single crystals. At low temperatures
two activation energies of impurities were estimated to 1 meV and 10
meV. The temperature dependence of the specific heat and the band
structure calculations provide also a semiconducting behaviour of
RuIn$_3$.
\end{abstract}

           

\maketitle

\section{\label{sec:level1}introduction}
Intermetallic compounds formed by elemental metals with good
conductivity are usually metallic conductors as well. 
A decade
ago, R. P\"ottgen \cite{Poettgen} in a study of the crystal structure
and physical properties of RuIn$_3$ reported on metallic conductivity
of this compound. However, for the isostructural compounds FeGa$_3$
and RuGa$_3$ electrical resistivity measurements revealed a
semiconducting behaviour \cite{Haeussermann}. Since isostructural
gallides and indides of the same group of transition elements have the
same number of conduction electrons, the same type of conduction
mechanism should be expected. All these compounds crystallise in the
tetragonal FeGa$_3$ structure (space group P4$_2$/mnm) and should show
anisotropic electrical properties.

The measurements on RuIn$_3$ \cite{Poettgen} were performed on
polycrystalline material obtained by solid state reactions in sealed
tantalum tubes, while the measurements on FeGa$_3$ and RuGa$_3$
\cite{Haeussermann} were performed on small unorientated single
crystals grown from a Ga-flux. In this article we report on
anisotropic transport properties and on caloric properties on RuIn$_3$
single crystals, which confirm the semiconducting nature of this
compound. The experimental results were strongly supported by LDA band
structure calculations, which give an indirect band gap of 0.41
eV. Furthermore, the band structure calculations provide a qualitative
explanation of the observed transport anisotropy.

\section{\label{sec:level2}experimental}
Single crystals of RuIn$_3$ were grown using the flux method \cite{Canfield} with Indium as reactant and as flux medium. An ingot of Ruthenium ($\cong$0.4 g) was placed at the bottom of a small Al$_2$O$_3$ crucible which was subsequently filled with liquid Indium. The mass ratio of Ru to In used was about 1:20. The crucibles were wrapped by Zr foil and placed in a vertical tube furnace under flowing Ar at atmospheric pressure. Afterwards the furnace was heated to 1150 $^\circ$C and cooled with a rate of 4 $^\circ$C/h to 600~$^\circ$C and below this temperature with a faster rate to 20 $^\circ$C.
\begin{figure}[htb]
\includegraphics[width=\columnwidth]{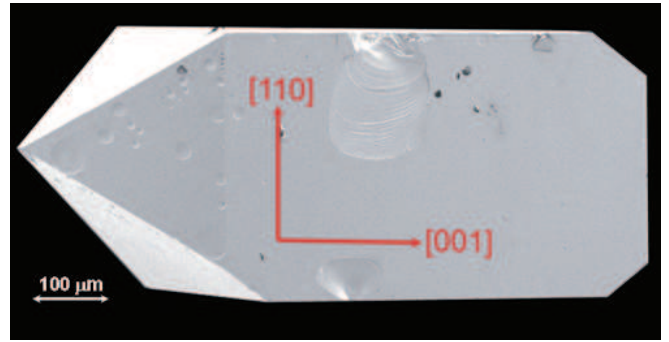}
\caption{\label{Fig1} SEM-picture of a well shaped RuIn$_3$ single crystal with [110]- and [111]-faces.}
\end{figure}
The crystals were removed from the solidified In by heating the crucibles to 200 $^\circ$C and decanting the liquid Indium excess. The single crystals were then etched in dilute HCl to remove the residual In from their surfaces. Well-shaped silvery-gray crystals with sizes of up to 5 mm and masses of about 250 mg were typically obtained. The crystals consist on parallelepipeds with [110]-faces and two pyramidal apexes with [111]-faces (see Fig. \ref{Fig1}). The different faces of the single crystals were characterised by x-ray diffraction and the orientations were controlled by Laue-diffraction. The stoichiometry of the crystals determined by EDX analysis gives a composition of 25$\pm$0.2 at\% Ru and 75$\pm$0.45 at\% In.\\
To measure the anisotropy of the resistivity the Montgomery-geometry \cite{Montgomery} was chosen (see inset Fig. \ref{Fig2}). The crystals were cut by spark erosion in cubes with edge lengths of 1.5 mm and supplied with four metallic contacts at the corners of the cubes. In addition measurements of the Hall-coefficient were performed on thin platelets of the same crystals. The measurements of the resistivity were done between 0.06 K and 400 K with a standard Lock-In technique. The measurements of the Hall-coefficient and specific heat were performed between 2 K and 300 K using a Quantum Design PPMS.

\section{\label{sec:level3}results}

\begin{figure}[htb]
\includegraphics[width=\columnwidth]{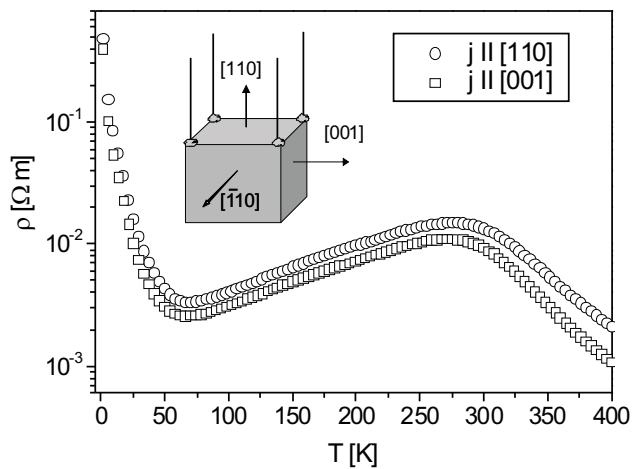}
\caption{\label{Fig2} Temperature dependence of the resistivity of RuIn$_3$ for different crystallographic directions obtained by the Montgomery-method. Inset shows a single crystal with contacts in Montgomery-geometry.}
\end{figure}

\begin{figure}[htb]
\includegraphics[width=\columnwidth]{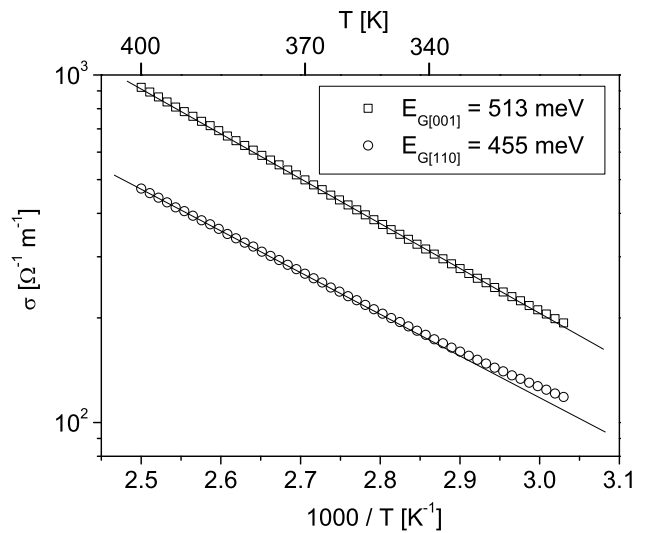}
\caption{\label{Fig3} Conductivity as a function of inverse temperature for different directions in the temperature range of the intrinsic conductivity.}
\end{figure}

\begin{figure}[htb]
\includegraphics[width=\columnwidth]{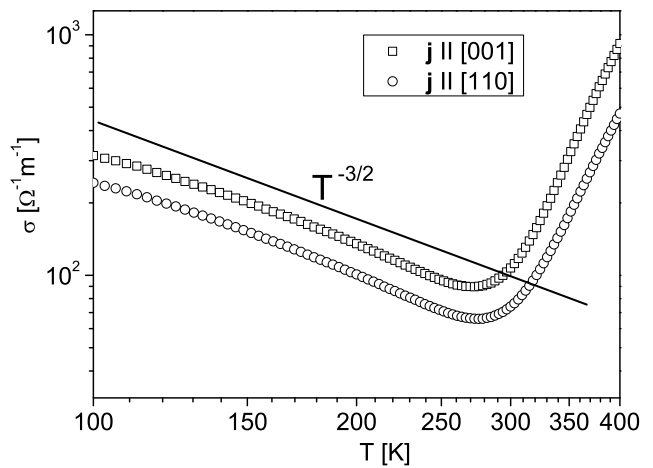}
\caption{\label{Fig4} Conductivity of RuIn$_3$ vs. temperature for both crystal orientations in a double-logarithmic plot together with a T$^{-3/2}$-dependence.}
\end{figure}

\begin{figure}[htb]
\includegraphics[width=\columnwidth]{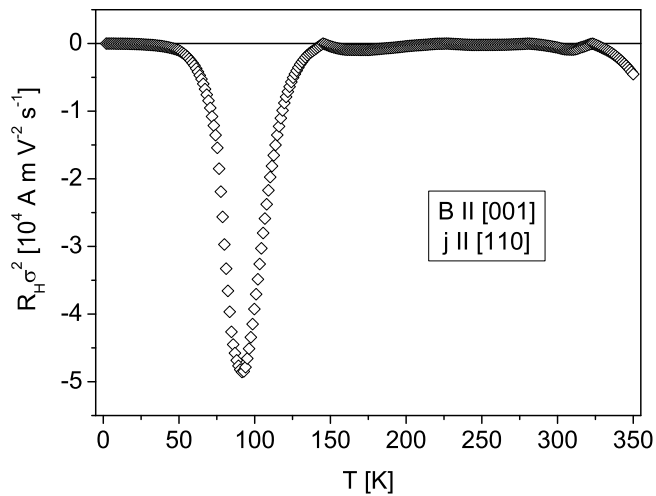}
\caption{\label{Fig5} The product $R_H(T)\sigma^2(T)$ as a function of temperature for the magnetic field \textbf{B}$\parallel$[001] and the current density \textbf{j}$\parallel$[110].}
\end{figure}

\begin{figure}[htb]
\includegraphics[width=\columnwidth]{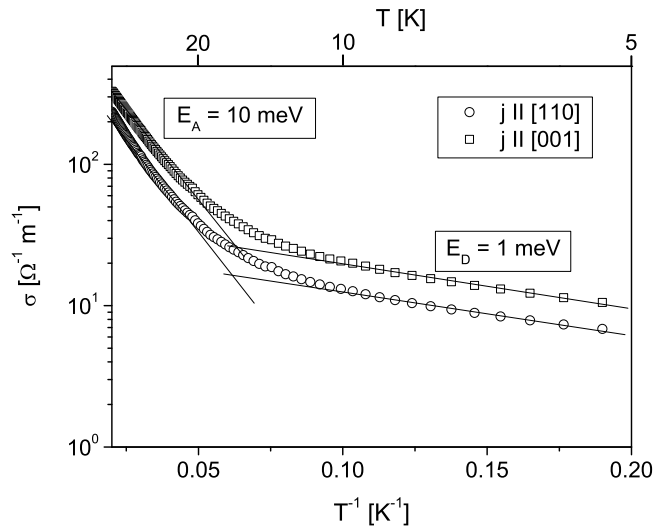}
\caption{\label{Fig6} Conductivity as a function of inverse temperature for different crystal orientations in the temperature range 5~K$\leq$T$\leq$50~K .}
\end{figure}

The temperature dependence of the resistivity obtained by the Montgomery-method on a single crystal of RuIn$_3$ is shown in Fig. \ref{Fig2}.
The resistivity ratio $\rho$[110]/$\rho$[001]=2.3 at T=400 K shows the anisotropy of the tetragonal crystals. The resistivity increases with decreasing temperature in the range from 400 K to 270 K. In this temperature range the conductivity $\sigma(T)$ can simply be fitted by the equation $\sigma(T)=\sigma_0\cdot\exp{(-E_G/2k_BT})$ for the intrinsic conductivity of a semiconductor. From a plot of $\ln{(\sigma/\sigma_0)}$ vs. inverse temperature shown in Fig. \ref{Fig3} the gap energy of RuIn$_3$ can be estimated to $E_{G[110]}$=0.455 eV  and  $E_{G[001]}$=0.513 eV, respectively.\\
In the temperature region between 270 K and 120 K the Hall-coefficient of all measured samples is more or less constant and therefore the carrier concentrations were assumed to be constant. In this case the temperature dependence of the conductivity is caused by the temperature dependence of the mobility. If the scattering of the carriers is dominated by acoustic phonons a $T^{-3/2}$ dependence should be observed. Fig. \ref{Fig4} shows the conductivities $\sigma(T)$  for both crystal orientations in a double-logarithmic plot together with a $T^{-3/2}$ straight line which demonstrates that in the range 120 K$\leq$T$\leq$250 K the scattering is dominated by acoustic phonons.\\
Below 120 K the Hall-coefficient of the RuIn$_3$ single crystals is negative and increases with decreasing temperature. In this temperature range the conductivity diminishes caused by the freezout of the extrinsic electrons and holes. Due to the different binding energies of the extrinsic carriers and the temperature dependence of the corresponding mobilities the temperature dependence of $R_H(T)$ is rather complex. In a two band model with electrons and holes the Hall-coefficient is given by
\begin{eqnarray}
R_H(T) = \frac{1}{e}\cdot\frac{n_h\mu^2_h-n_e\mu^2_e}{(n_h\mu_h+n_e\mu_e)^2}
\label{eq:one}
\end{eqnarray} 
with the concentrations $n_h, n_e$ and the mobilities $\mu_h, \mu_e$ of the holes and the electrons. From the measured conductivity $\sigma(T)$  and the Hall-coefficient $R_H(T)$ the product
\begin{eqnarray}
R_H(T)\cdot\sigma^2(T) = e\cdot(n_h\mu^2_h-n_e\mu^2_e)
\label{eq:two}
\end{eqnarray} 			      
can be obtained, which temperature dependence is easier to interpret. In Fig. \ref{Fig5} the product $R_H(T)\sigma^2(T)$ is given as a function of temperature. At low temperatures (T$<$20 K) the value of $\left|R_H(T)\sigma^2(T)\right|$ is very small since most of the carriers are frozen at their impurity sites. $\left|R_H(T)\sigma^2(T)\right|$ steeply increases above 50 K reaching a maximum at about 90 K. Since the sign of $R_H(T)\sigma^2(T)$ is negative the increase is caused by the thermal activation of electrons from donors with a small activation energy $E_D$.
Above 90 K $\left|R_H(T)\sigma^2(T)\right|$ decreases strongly reaching $R_H(T)\sigma^2(T)$=0  at about 150 K. The decrease of $\left|R_H(T)\sigma^2(T)\right|$   in the range 90 K$\leq$T$\leq$150 K will be produced by the thermal activation of holes from acceptors with a larger activation energy $E_A$. It should be noted that $\left|R_H(T)\sigma^2(T)\right|$ is comparably small in the whole temperature range 150 K$\leq$T$\leq$300 K. This could be a hint to the fact, that both kinds of carriers originate from the same type of structural defects. The activation energies $E_D$ and $E_A$ can be obtained from the plot of $\ln{(\sigma/\sigma_0)}$ vs. inverse temperature $T^{-1}$ in the temperature range 5 K$\leq$T$\leq$50 K , which is given in Fig.~\ref{Fig6}.
Typical values obtained for the activation energies on different
single crystals are $E_D$=1.0-1.5 meV  for the donors and $E_A$=10meV
for the acceptors.
In the temperature range $T>320$K the quantity
$R_H(T)\cdot\sigma^2(T)$ becomes negative again. In this regime
intrinsic conductivity dominates and $n_h=n_e=n_i$. From
$R_H\cdot\sigma^2=en_i(\mu_h^2-\mu_e^2)<0$ it follows $\mu_e>\mu_h$,
which in turn implies $m^\ast_e<m^\ast_h$. The interpretation of this
observation will be given in connection with band structure calculations.

\begin{figure}[htb]
\includegraphics[width=\columnwidth]{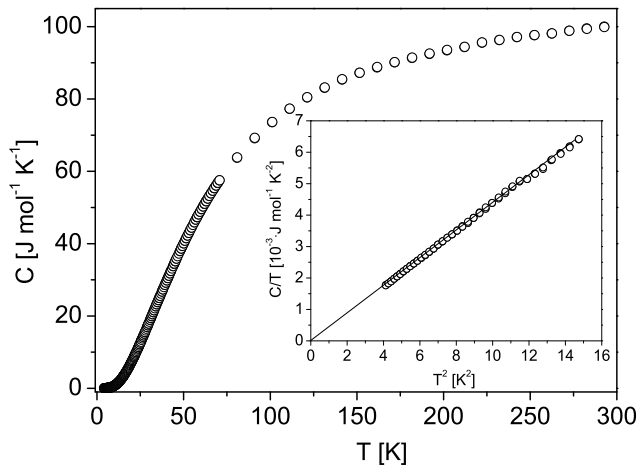}
\caption{\label{Fig7} Specific heat of RuIn$_3$ in the temperature range 2~K$\leq$T$\leq$300~K. Inset: Plot of C/T vs. T$^2$ in the low temperature range  2~K$\leq$T$\leq$4~K.}
\end{figure}

The specific heat of RuIn$_3$, measured on a single crystal with a
mass of m$\cong$25 mg is given in Fig. \ref{Fig7}. $C(T)$ shows the
typical temperature dependence of a semiconductor or an insulator in
the temperature range 2~K$\leq$T$\leq$300~K  with a value of 100 J
mol$^{-1}$ K$^{-1}$ at 300 K. The inset of Fig. \ref{Fig7} shows the
low temperature data (2~K$\leq$T$\leq$4~K) in a C/T vs. T$^2$  plot.  
The extrapolation to T=0 K goes through the origin, and indicates that the contribution of the carriers is very small. This fact corroborates the semiconducting behaviour of our transport measurements. The value of the slope is A=4.5$\cdot$10$^{-4}$~J~mol$^{-1}$~K$^{-4}$. From this result the Debye-temperature $\Theta_D$=$\left(\frac{12\pi^4Nk_B}{5A}\right)^{1/3}$ of RuIn$_3$ can be calculated to $\Theta_D$=258 K.

\section{\label{sec:level4}band structure calculation}

\begin{figure}[htb]
\includegraphics[width=\columnwidth]{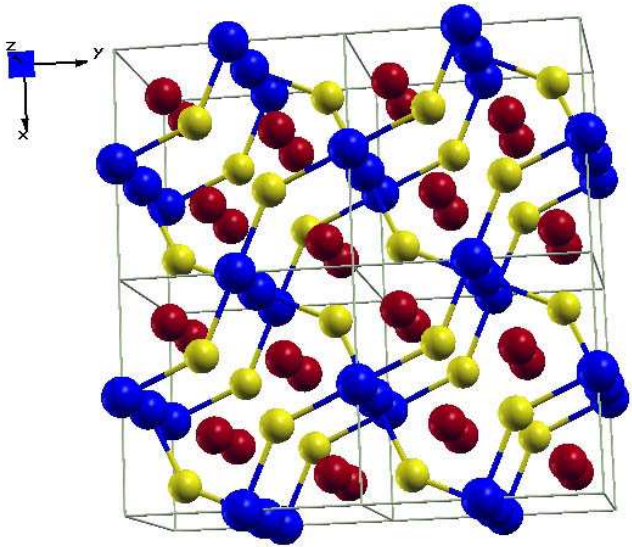}
\caption{\label{Fig8} Crystal structure of RuIn$_3$. Red-In1 sites, blue-In2 sites, yellow-Ru sites.}
\end{figure}

State-of-the-art electronic structure calculations from first
principles are based on density functional theory (DFT) within the local
density approximation (LDA). To calculate the electronic structure of 
RuIn$_3$, the TB-LMTO-ASA package
v.47 (Tight Binding, Linear Muffin-Tin Orbitals, Atomic Sphere
Approximation) \cite{Andersen} was used with experimentally obtained
values of the lattice constants 
($a=b=7.003$~\AA, $c=7.247$~\AA, tetragonal space
group P4$_2$/mnm $\#$136 \cite{Poettgen}). The Ru sites occupy 4f Wyckoff positions
($x=0.3451$, $y=0.3451$, $z=0$) and there exist two crystallographically inequivalent In
sites at positions 4c for In1 ($x=0$, $y=0.5$, $z=0$) and 8j for In2
($x=0.15547$, $y=0.15547$, $z=0.26224$). The crystal
structure is presented in Fig. \ref{Fig8}.
Atomic spheres radii were chosen to be R(Ru)=2.88 a.u., R(In1)=2.92
a.u. and R(In2)=2.97 a.u.. The orbital basis consists of 5s, 5p, 4d
muffin-tin orbitals for Ru and 5s, 5p for In1 and In2 sites. To fill
the space of the unit cell several empty (without nuclei charge)
atomic spheres were introduced. The calculations were performed with
405 irreducible \textbf{k}-points (16$\times$16$\times$16 spacing) in
the first Brillouin zone.

\begin{figure}[htb]
\includegraphics[width=\columnwidth]{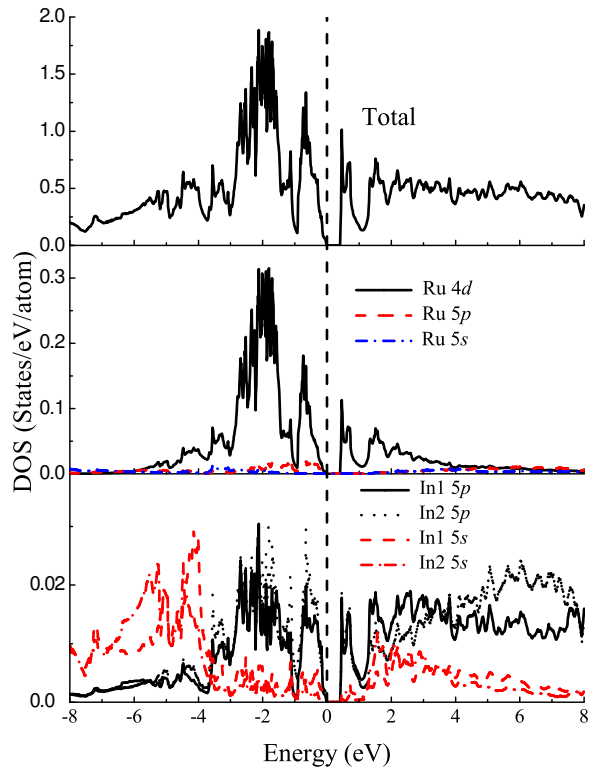}
\caption{\label{Fig9} LDA calculated densities of states for RuIn$_3$. The upper panel shows total DOS. The middle panel displays Ru 4d (full line), Ru 5p (dashed line) and Ru 5s (dotted line) states. The lower panel contains In1 5p (full line), In2 5p (long dashed line), In1 5s (dotted line) and In2 5s (dashed line) states. Fermi level corresponds to zero.}
\end{figure}

In Fig. \ref{Fig9} we present the densities of states (DOS) obtained
within the LMTO method. The upper panel displays the total DOS with an
energy gap of 0.41 eV. The individual contributions of Ru 5s, 5p and
4d states and the 5s, 5p states of both In sites to the total DOS are
shown in the middle and lower panels of Fig.~\ref{Fig9}, respectively.
By comparing the total and partial DOS, 
one can identify several evident features. The energy range below
-4.5~eV is dominated by In1(2) 5s states. In the energy interval
$-4.5\ldots2$ eV one observes that the 5p states of In1 and In2
hybridise relatively strongly with Ru 4d states. Finally, above 2 eV 
there are predominantly In 5p states (see lower panel,
Fig.~\ref{Fig9}).

\begin{figure}[htb]
\includegraphics[width=\columnwidth]{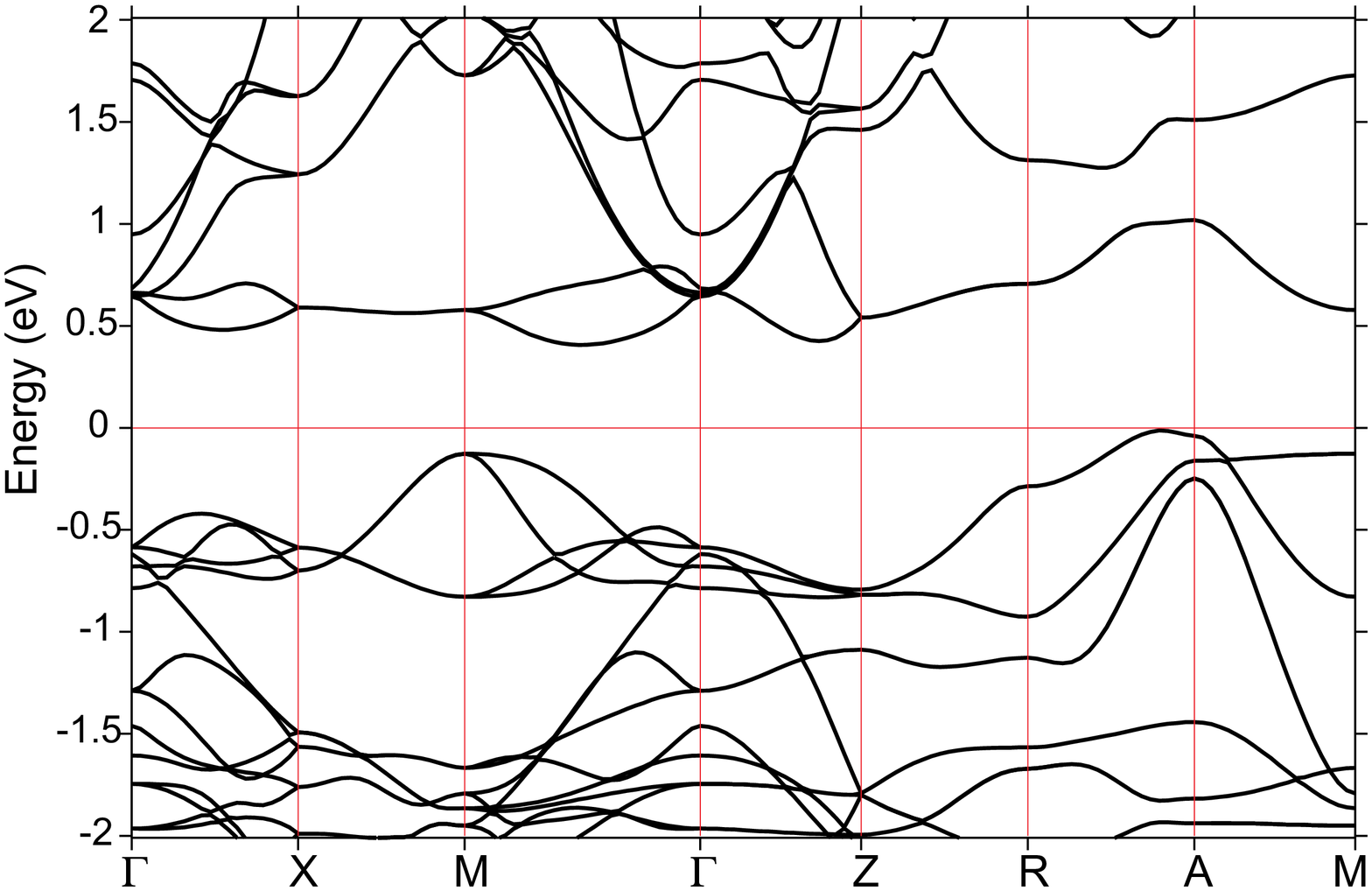}
\caption{\label{Fig10} Band structure of RuIn$_3$ in the vicinity of
  the Fermi level calculated within standard DFT/LDA. The Fermi level corresponds to zero.}
\end{figure}

The observed anisotropy of transport properties with different
resistivities along $ab$ and $c$ directions is quite generally
consistent with the simple tetragonal crystal structure of the
compound. In order to qualitatively explain why
the $ab$ resistivity is larger than the one along $c$
we first note that 
the properties of hole transport
will be determined by the top of the
valence band located at \textbf{k}-point $(0.48, 0.48, 0.48)$ of the
RA high symmetry direction (see Fig. \ref{Fig10}),
because the energy gap is much larger than the experimental temperature
and
there are no other bands below this one within the interval of
energies corresponding to the experimental temperature.

From the LDA band structure analyses one can conclude
that the orbitals contributing to the 
top of the valence band
along the RA high symmetry direction in Fig. \ref{Fig10}
can be separated into
those lying in the XY plane and those along the Z-axis.
The former are
superpositions of In2 5p$_x$, 5p$_y$ and Ru 4d$_{xz}$, 4d$_{yz}$ orbitals,
the latter obtained from In1 5p$_z$ and again Ru 4d$_{xz}$, 4d$_{yz}$
orbitals,
and a small In2 5p$_z$ contribution
\cite{seven}. 
There is also an In1 5s contribution to the top of the
valence band which is anisotropic. Contributions of other
symmetries are negligible.

The strongest orbital overlap
is obtained for 
nearest neighbours.
In the XY-plane these nearest neighbours are In2-Ru and Ru-Ru,
In1 sites do not have nearest neighbours in the XY-plane.
Along the Z-axis nearest neighbours are Ru-In1 and In1-In2.
Together with the previous classification of the orbitals
contributing to the top of the valence band
we can have 
In2-Ru $pd\pi$ hybridisation (In2 5p$_z$ and Ru 4d$_{xz}$ and 4d$_{yz}$)
and Ru-Ru $dd\delta$ hybridisation (Ru 4d$_{xz}$-4d$_{xz}$, Ru 4d$_{yz}$-4d$_{yz}$) \cite{Koster} in the XY-plane.
These two types of hybridisation are known to be quite weak
\cite{Andersen78}.
Along the Z-axis the possible hybridisation are Ru-In1 $sd\sigma$
(Ru 4d$_{xz}$,4d$_{yz}$ and In1 5s), In1-In2 $sp\sigma$
(In1 5s and In2 5p$_x$,5p$_y$,5p$_z$) and
$pp\sigma$ (In1 5p$_z$ and In2 5p$_z$),
which in principle are stronger \cite{Andersen78} than the ones 
found
for the XY-plane, thus qualitatively explaining 
the better conductivity along the Z-axis.

A similar argumentation can be applied to the electron transport dominated 
at low temperatures
by the bottom of the conduction band in the middle of the $\Gamma M$ direction.

Since the angles between bonds are not ideal in RuIn$_3$
(90$^\circ$ or 180$^\circ$), there will of course
be other types of hybridisation. Furthermore, next nearest neighbour
hoppings will contribute as well. All these effects will
further reduce the anisotropy and
lead to three dimensional behaviour despite the layered structure 
of RuIn$_3$.
Given all these partially compensating effects a final quantitative
answer on the size of the anisotropy thus surely requires more
detailed numerical investigations.

The band structure of RuIn$3$ given in Fig.\ \ref{Fig10} shows a maximum of the
valence band along $RA$ in the Brillouin zone near $A$. The calculated effective
hole mass is $m_h\approx1.2m_0$. The minimum of the conduction band is formed by
a flat band along $\Gamma M$ with a large effective electron mass $m_{e1}\approx2.6m_0$.
Therefore in the intrinsic regime the quantity $R_H\cdot\sigma^2$ should be
positive in contrast to the observed high temperature behaviour in
Fig.\ \ref{Fig5}.  However, the band structure shows a second minimum of the
conduction band along $\Gamma Z$ near $Z$ with a substantial smaller effective
mass of $m_{e2}\approx1.1m_0$. The energy difference between the two conduction band minima is only $20$meV, which is smaller as the thermal energy $k_{\rm B}T$
at $T\ge300$K. Hence, the small effective electron mass $m_{e2}$ of the
conduction band minimum near $Z$ with the implication $\mu_{e2}>\mu_h$ is
responsible for the negative sign of $R_H\cdot\sigma^2$ observed by the
experiments in the intrinsic range.
\section{Summary\label{sec:summary}}
Measurements of the resistivity and of the Hall-coefficient of RuIn$_3$ single
crystals identify this compound as a semiconductor with a gap energy of 
$E_G=0.46$-$0.51$eV, in contrast to previously published results, which report
on metallic conductivity. The calculated band structure based on the density
functional theory gives a similar value of $E_G=0.41$eV.

The experimentally observed slight anisotropy of the resistivity
$\rho[110]>\rho[001]$ can qualitatively be explained by a theoretical symmetry
analyses which predict a better conductivity along the $c$-axis of the tetragonal
compound. The negative sign of the Hall-coefficient in the intrinsic region can
also be explained by the calculated conduction band structure with two
energetically nearly equivalent minima but with very different effective
electron masses.

In the extrinsic range at low temperatures a small activation energy for
electrons from donors of $E_D=1.0$-$1.5$meV and a larger activation energy for
holes from acceptors of $E_A=10$meV were determined from the transport
measurements. Both kinds of extrinsic carriers seems to originate from the same
type of structural defects, since $R_H\cdot\sigma^2$ is nearly zero in a large temperature range
$150$K$<T<320$K. The low temperature data of the specific heat show a negligible
contribution of charge carriers and corroborate the semiconducting nature of 
RuIn$_3$.

\begin{acknowledgments}
IN thanks A. Postnikov for many helpful discussions.
This work was supported in part by RFBR grants 05-02-16301 (IN),
05-02-17244 (IN), 06-02-90537 (IN), by the joint UrO-SO project
(IN), and programs of the Presidium of the Russian Academy of
Sciences (RAS) "Quantum macrophysics" and of the Division of Physical
Sciences of the RAS "Strongly correlated electrons in semiconductors,
metals, superconductors and magnetic materials" and the by the DFG
through the collaborative research grant SFB 602 (TP,IN). I.N. acknowledges
support from the Dynasty Foundation and International Center for
Fundamental Physics in Moscow program for young scientists and also
from the grant of President of Russian Federation for young PhD
MK-2242.2007.02, and thanks the Faculty of Physics of the Georg-August
University of G\"ottingen for its hospitality.
\end{acknowledgments}


\begin{thebibliography}{9}
\bibitem{seven} Since RuIn$_3$ cristallizes in a tetragonal structure
  without distortions, there is no need to introduce a local
  coordinate system for each particular atom to properly resolve the
  orbitals with respect to their magnetic quantum number.
\end{thebibliography}
\end{document}